\documentclass[sigconf]{acmart}
\AtBeginDocument{%
  \providecommand\BibTeX{{%
    \normalfont B\kern-0.5em{\scshape i\kern-0.25em b}\kern-0.8em\TeX}}}

\usepackage{soul, color, xcolor}
\usepackage{algorithm}
\usepackage{algorithmic}

\usepackage{soul}

\newcommand\ie{i.e.}
\newcommand\eg{e.g.}

\acmConference[MM ’21]{Proceedings of the 29th ACM International Conference on Multimedia (MM ’21)}{October 20--24th, 2021}{Chengdu, Sichuan, China}
\acmBooktitle{Proceedings of the 29th ACM International Conference on Multimedia (MM ’21),
  October 20--24th, 2021, Chengdu, Sichuan, China}

\begin{document}

\title{Multi-trends Enhanced Dynamic Micro-video Recommendation}


\author{Yujie Lu}
\affiliation{%
  \institution{UC Santa Barbara}
  \country{United States}
}
  
\email{yujielu@zju.edu.cn}

\author{Yingxuan Huang}
\affiliation{%
  \institution{The University of Hong Kong}
  \country{China}
}
\email{eloise@connect.hku.hk}

\author{Shengyu Zhang}
\authornotemark[1]
\affiliation{%
  \institution{Zhejiang University}
  \country{China}
}
\email{sy\_zhang@zju.edu.cn}

\author{Wei Han}
\affiliation{%
  \institution{Singapore University of Technology and Design}
  \country{China}
}
  
\email{henryhan88888@gmail.com}

\author{Hui Chen}
\affiliation{%
  \institution{Singapore University of Technology and Design}
  \country{China}
}
\email{hui_chen@mymail.sutd.edu.sg}

\author{Fei Wu}
\authornotemark[1]
\affiliation{%
  \institution{Zhejiang University}
  \country{China}
}
\email{wufei@zju.edu.cn}
\author{Zhou Zhao}
\authornote{Corresponding Authors.}
\affiliation{%
  \institution{Zhejiang University}
  \country{China}
}
\email{zhaozhou@zju.edu.cn}

\renewcommand{\shortauthors}{Yujie Lu, et al.}

\begin{abstract}

The explosively generated micro-videos on content sharing platforms call for recommender systems to permit personalized micro-video discovery with ease. Recent advances in micro-video recommendation have achieved remarkable performance in mining users' current preference based on historical behaviors. However, most of them neglect the dynamic and time-evolving nature of users' preference, and the prediction on future micro-videos with historically mined preference may deteriorate the effectiveness of recommender systems. In this paper, we propose to explicitly model dynamic multi-trends of users' current preference and make predictions based on both the history and future potential trends. We devise the DMR framework, which comprises: 1) the implicit user network module which identifies sequence fragments from other users with similar interests and extracts the sequence fragments that are chronologically behind the identified fragments; 2) the multi-trend routing module which assigns each extracted sequence fragment into a trend group and update the corresponding trend vector; 3) the history-future trend prediction module jointly uses the history preference vectors and future trend vectors to yield the final click-through-rate. We validate the effectiveness of the proposed framework over multiple state-of-the-art micro-video recommenders on two publicly available real-world datasets. Relatively extensive analysis further demonstrate the superiority of modeling dynamic multi-trend for micro-video recommendation.

\end{abstract}




\maketitle
\pagestyle{plain}
\section{Introduction}
In recent years, the amount of searchable micro-videos has increased dramatically and exacerbated the need for recommender systems that can effectively mine users' preference and identify potentially interested micro-videos in a personalized manner. Due to the powerful representation learning capacity, the rapid development of deep learning techniques has nourished the research field of recommendation \cite{Lu_Zhang_Huang_Wang_Yu_Zhao_Wu_2020,Du_Wang_He_Li_Tang_Chua_2020,He_Deng_Wang_Li_Zhang_Wang_2020,Kang_McAuley_2018,Li_Xu_Jiang_Cao_Huang_2020,Sun_Liu_Wu_Pei_Lin_Ou_Jiang_2019,Tang_Wang_2018,Wang_Huang_Zhao_Zhang_Zhao_Lee_2018,Wang_He_Wang_Feng_Chua_2019,Wei_Wang_Nie_He_Chua_2020,Wei_Wang_Nie_He_Hong_Chua_2019,Wu_Tang_Zhu_Wang_Xie_Tan_2019,Yang_Xie_Wang_Yuan_Ding_Yan_2020,Yu_Gan_Wei_Cheng_Nie_2020}. Such a development also gives rise to diverse models for video recommendation, which can be roughly categorized to collaborative filtering \cite{Baluja_Seth_Sivakumar_Jing_Yagnik_Kumar_Ravichandran_Aly_2008,Huang_Cui_Jiang_Hong_Zhang_Xie_2016}, content-based filtering \cite{Cui_Wang_Su_2014,Mei_Yang_Hua_Li_2011,park2010online,Zhou_Chen_Zhang_Cao_Huang_Wang_2015,Dong_Li_Xu_Yang_Wang_2018}, and hybrid ones \cite{Chen_Wang_Huang_Mei_2012,Chen_Song_Nie_Wang_Zhang_Chua_2016,Yan_Sang_Xu_2015}.


Compared with professional video recommendation, micro-video recommendation poses many unique challenges. For example, micro-videos typically lack of meta-data (\eg, genre, director, actor/actress, which are commonly available in professional videos), leading to semantic gap in representation \cite{Chen_Liu_Zha_Zhou_Xiong_Li_2018}. Moreover, users might be interested in multiple topics of videos simultaneously, \ie, diverse interests, and yield interests to different extends (\eg, like, follow, click), \ie, multi-level interests \cite{Li_Liu_Yin_Cui_Xu_Nie_2019}. Recent years have witnessed much progress to confront the above challenges in this vein. THACIL~\cite{Chen_Liu_Zha_Zhou_Xiong_Li_2018} employs temporal block splitting and hierarchical multi-head attention to model diverse interests across blocks. ALPINE~\cite{Li_Liu_Yin_Cui_Xu_Nie_2019} models users' dynamic interests by constructing temporal behavior graph and devising the temporal graph-based LSTM. MTIN~\cite{Jiang_Wang_Wei_Gao_Wang_Nie_2020} considers personalized importance decay over time and diverse interests using item-level temporal mask and group routing mechanism, individually. In spite of the great advances of these works, we argue that solely modeling the historical behaviors deteriorates the capacity of user modeling capturing \textit{diverse} and \textit{dynamic} users' interests. For example, MTIN \cite{Jiang_Wang_Wei_Gao_Wang_Nie_2020} assigns historically interacted items to one of six interest groups and accordingly updates the six interest vectors. Since users' interests are by nature dynamic, the interests learned from the logged data might be out-of-date or at least limited to the history, falling short to recommend fresh items and hurting the recommendation diversity. Therefore, capturing dynamic interest trends based on (but not limited to) historical items can be an indispensable function for high-quality recommender systems.

Towards this end, we devise the multi-trends framework for dynamic micro-video recommendation, abbreviated as DMR. We start from the perspective that trends refer to the possible future directions of the current interest implied by the logged interactions. Since we have no access to items interacted in the future, DMR encapsulates an implicit user network construction module that first identifies sequence fragments that yield similar interests as the current sequence from similar users. Then, we constructs possible trending sequences by extracting the sequence fragments that are chronologically behind the identified ones. We note that some trending sequences may share similar interests and representing each sequence as an individual interest may introduce unnecessary noises and computation costs. Towards this end, inspired by \cite{Jiang_Wang_Wei_Gao_Wang_Nie_2020,Li_Liu_Wu_Xu_Zhao_Huang_Kang_Chen_Li_Lee_2019}, we devise a multi-trend routing module that transforms multiple trending sequences to fewer number of multiple trend interest vectors. However, extracting trending sequences and mapping them to trend vectors for each testing inference might hurt the serving efficiency of industrial deployment. Thus, multi-trend routing module constructs a fixed-length trend memory for each user and read-writes the memory during training. For memory read-writing, we propose to assign trending sequences to memory slots in a soft way and power the process with attention mechanisms. During inference, we directly take the off-the-shelf history/trending vectors without extracting or transforming trending sequences, and thus addressing the efficiency issue. Predictions are performed with the history-trend joint prediction module.

To this end, DMR framework makes predictions based on both the history interests implied by the historical behaviors as well as multi-trends implied in similar users, which helps to capture even more diverse and dynamic interests compared with existing micro-video recommenders. We validate the effectiveness of DMR on micro-video recommendation benchmarks. The substantial improvement over state-of-the-art comparison methods and in-depth model analysis demonstrate the superiority of modeling multi-trend for micro-video recommendation. Overall, this paper has the following contributions:

\begin{itemize}
	\item We propose to capture even more diverse and dynamic interests beyond the historical behaviors by modeling the possible interest trends for micro-video recommendation.
	\item We devise the novel DMR framework that encapsulates the implicit user network construction module, which extracts trending sequences from similar users, the multi-trend routing module, which performs dynamic trending memory read-write and improves the inference efficiency at the inference stage, and the history-trend joint prediction module.
	\item We conduct extensive experiments on micro-video recommendation benchmarks, of which the results show DMR framework achieves high-quality recommendation with improvement on both accuracy and diversity.
\end{itemize}

\section{Related Work}
\label{sec:related}

\subsection{Video Recommendation}
The methods for recommendation can be generally classified into two categories. 
Early algebraic approaches adopted collaborative filtering~\cite{konstan1997grouplens, ding2005time, he2017neural, sarwar2001itemCF} or model-based methods~\cite{rendle2010factorizing, deshpande2004item, wang2015learning, kim2014twilite} to estimate user-item correlations and make predictions about users' future interests. Collaborative filtering (CF) assumes that users sharing the same opinion on one issue tend to have more similar opinions on other issues~\cite{koren2015advances}, and thus it makes predictions specific to each user through information gleaned from other users~\cite{terveen2001beyond}. 
Due to the extreme high computational complexity and data sparsity in traditional CF~\cite{deshpande2004item, papagelis2005alleviating}, model-based methods alleviate this overhead by mapping user-item interactions into matrix entries, then apply factorization to the characteristic matrix to build nonlinear models that estimate correlations (i.e. preferences) between every pair of user and item and employ Hidden Markov models (HMM) to capture temporal trends of preferences~\cite{rendle2010factorizing}.

Recently, as the major advances in deep learning techniques, a wealth of research has sprung up on incorporating them into recommender systems. Most of work reformulated traditional estimation problem as learning task based on deep neural networks~\cite{fan2019graph, smirnova2017contextual,he2018nais}. In the field of video recommendation, representative work focuses on content-based learning~\cite{paul2016youtubednn, wang2019overview, wei2019neural, deldjoo2016content}, in which features of videos are extracted into embedding vectors and then matched with user representations that indicate individual preference. To name a few, \citet{chen2017attentive} tackled the item- and
component-level implicit feedback issue in multimedia recommendation by learning independent video and users characteristics in a unified hierarchical attention network and then reckoning pair-wise scores as a measure of user preference. Although these works improve the accuracy of user modeling, they lack a clear partition of history and future for the given dataset and hence may encounter prediction bias due to mixing the two parts together. Our work adopts a multi-step time partition and similarity matching approach to alleviate this issue.

\subsection{User Behavior Modeling}
Modeling latent user interest from historical behaviors is commonly used in recommender systems. In the past two decades, a variety of approaches have been proposed, ranging from Markov chains~\cite{shani2005mdp,rendle2010factorizing,he2016Vista,he2016fusing,norris1998markov} and traditional collaborative filtering~\cite{koren2009collaborative,ding2005time,salakhutdinov2008probabilistic} to deep representation learning~\cite{qu2016product,zhou2018deep}. The approaches based on Markov decision processes implicitly track user state dynamics to predict future behaviors. For example, \citet{rendle2010factorizing} captured long-term user interest via personalized transition graphs over underlying Markov chains. \citet{he2016fusing} integrated similarity-based methods with Markov chains smoothly in personalized sequential recommender systems. Besides, temporal collaborative filtering is proposed to deal with the drifting user preferences. \citet{koren2009collaborative} offered a paradigm that tracks time changing behaviors throughout the life span of the data. 

With the development of deep learning, more and more researchers adopted deep neural networks (DNN) to model the user dynamics in recommender systems. Particularly, \citet{hidasi2015gru4rec} applied recurrent neural networks (RNN) to model the whole session and introduced a new ranking loss function to make recommendations more accurate.  \citet{tang2018personalized} utilized convolutional filters to embed a sequence of recent items into an "image" in the time and latent spaces as well as learn sequential patterns as local features of the image. \citet{wu2019session} considered session sequences as graph structured data and used graph neural networks (GNN) to capture complex transitions of items. Recently,  self-attention mechanism~\cite{vaswani2017attention} has been widely employed in recommender systems\cite{kang2018self}. For instance, \citet{wu2020deja} proposed a Contextualized Temporal Attention Mechanism to weigh historical actions’ influence on not only what action it is, but also when and how the action took place.

However, previous work does not consider the influence of future information when modeling user behaviors in history sequences. In this work, we constructed a user-item heterogeneous graph to capture future interactions of each user with items.

\section{Methodology}
\label{sec:methodology}

\begin{figure*}
    \centering
    \includegraphics[width=\textwidth]{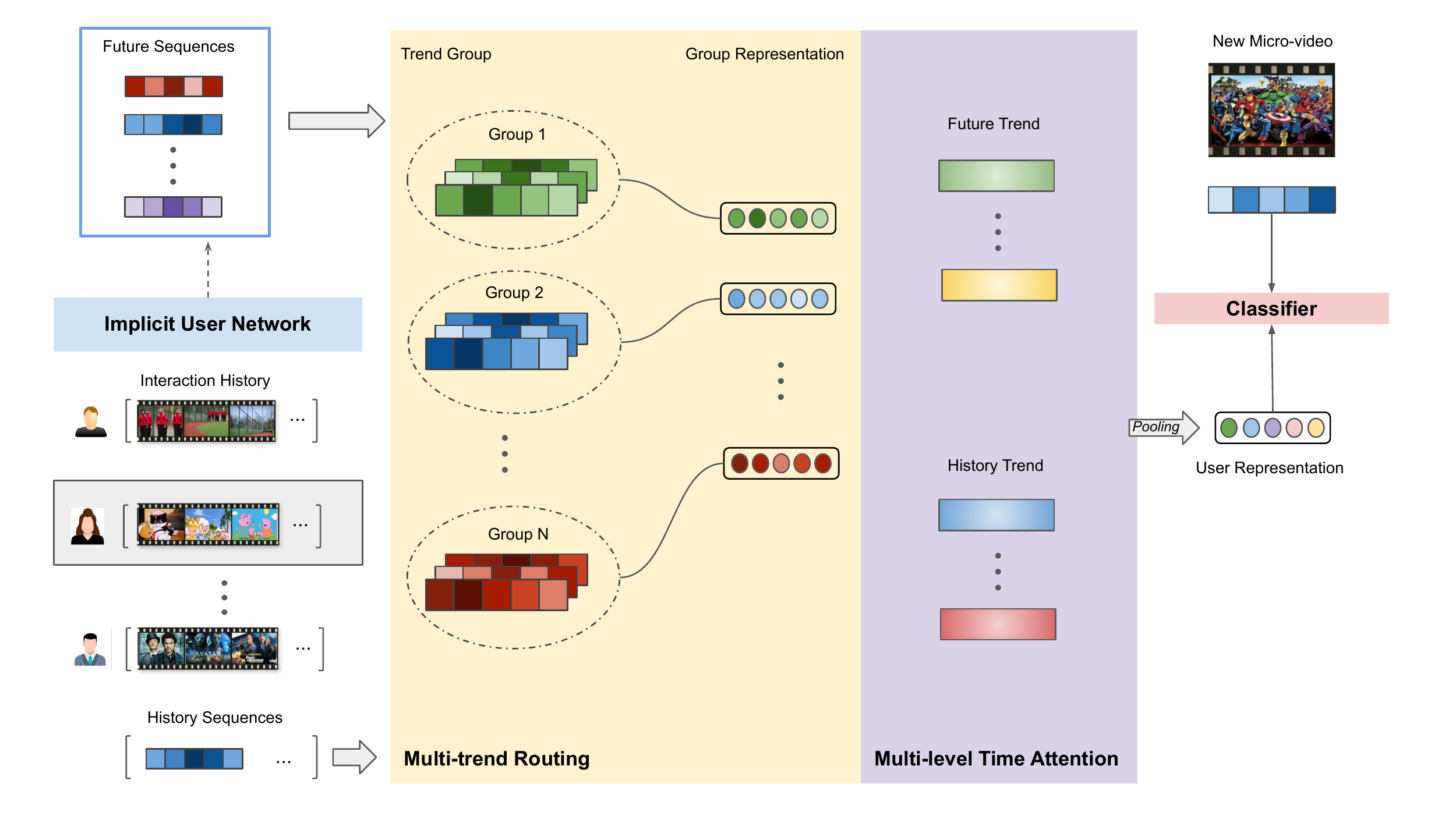}
    \caption{Network Architecture of DMR. DMR is composed of an implicit user network module, a multi-trend routing module, a multi-level time attention layer and a prediction layer. Based on the users' historical interactions, we build a implicit user network to construct future sequences. The multi-trend module are applied on the current user's history sequences and future sequences in parallel to get representation of each trend group. The multi-level time attention mechanism are applied before the pooling layer to generate the history trend representation and future trend representation, which is further concatenated as dynamic user preference representation. Finally, the user representation and the candidate micro-video embedding are utilized for prediction in the classifier.}
    \label{fig:framework}
\end{figure*}

In this section, we first formulate the micro-video recommendation problem, and then introduce the proposed framework in detail. As illustrated in Figure~\ref{fig:framework}, our proposed DMR framework for dynamic micro-video recommendation mainly comprises of three modules:1) Pearson Correlation Coefficient enhanced implicit user network module; 2) A history-future multi-trend joint routing module; 3) A multi-level time-aware attention module.

\subsection{Problem Formulation}
In a typical micro-video recommendation scenario, we have a set of users and a set of micro-videos, which can be denoted as ${U=\{u_1, u_2, u_3, ..., u_{|U|}\}}$ and ${V=\{v_1, v_2, v_3, ..., v_{|V|}\}}$ respectively. Let ${I_u = \{ x_1^u, x_2^u, ..., I_{|I_u|}^u \}}$ represent the sequence of interacted micro-videos ${x \in I_u}$ of user ${u \in U}$, which is sorted in a chronological order according to the timestamp of each interaction, and ${x_t^u}$ denote the micro-video that the user ${u}$ has interacted with at timestamp ${t}$. The interaction sequence ${I_u}$ is split into ${I_{+}}$ and ${I_{-}}$ which represent the micro-videos clicked by the user and the ones not clicked respectively. Given the user's historical micro-video interaction behaviors, the investigated goal of the micro-video recommendation task in this paper is to predict the probability that the new candidate micro-video will be clicked by user ${u}$. Notations are summarized in Table~\ref{tab:notation}.

\begin{table}[h]
\caption{Notations.}
\centering
\resizebox{\columnwidth}{!}{%
\begin{tabular}{l l}
\toprule
Notation & Description \\

    \midrule
    u   & a user \\
    v   & a micro-video \\ 
    x   & an interaction \\ 
    d   & the dimension of user/micro-video embeddings \\
    t   & the number of trends \\
    U   & the set of users \\
    V   & the set of micro-videos \\
    I   & the set of interactions \\
    T   & the trends set \\
    \bottomrule
\end{tabular}%
}
    \label{tab:notation}
\end{table}

Specifically, each instance is represented by a tuple ${(I_u, A_i)}$, where ${I_u}$ denotes the set of items interacted by user ${u}$, ${A_i}$ the features of target item ${i}$ including the information of interaction timestamp and micro-video embeddings. Through implicit user network module, we extract relative future sequence of user ${u}$ based on ${I_u}$ and their similar users' historical interaction ${I_{u'}, u' \in U}$. The detail will be illustrated in Section~\ref{sec:userNetwork}.

To model diverse user preferences dynamically, DMR learns a function ${f}$ for mapping history trend set ${T_u^h}$ and future trend set ${T_u^f}$ into user representations, which can be formulated as:
\begin{equation}
    \overrightarrow{e_u} = f(T_u^h, T_u^f)
\label{eq:userRepFunc}
\end{equation}
where ${\overrightarrow{e_u} \in \mathbb{R}^{d\times1}}$ denotes the representation vector of user ${u}$, ${d}$ the dimension. Besides, the representation vector of target micro-video ${i}$ is obtained by an embedding function ${g}$ as:
\begin{equation}
    \overrightarrow{e_i} = g(A_i)
\label{eq:itemRepFunc}    
\end{equation}
where ${\overrightarrow{e_i} \in \mathbb{R}^{d\times1}}$ denotes the representation vector of target micro-video ${i}$.

Based on the learned user representation vector and micro-video representation vector, the probability of candidate micro-video is calculated using the likelihood function ${P}$ as:
\begin{equation}
    p(i|U, V, X) = P(\overrightarrow{e_u}, \overrightarrow{e_i})
\label{eq:pFuncDemo}
\end{equation}
where ${\overrightarrow{e_i}}$ is the embedding of target item ${i}$ from set of micro-videos ${V}$. Our framework outputs the click probabilities of the candidate micro-video to rank the personalized recommendation list. Then the system provide precise and diversified recommendation for each user, which entails potential preference of the specific user as they are most likely to interact with the recommended micro-videos.

The objective function for training our model is described in Section~\ref{sec:prediction} We use the Adam optimizer to train our method.

\subsection{Overview}
The overall structure of our proposed framework DMR is illustrated in Figure~\ref{fig:framework}, which is composed of an implicit user network module, a multi-trend routing module, a multi-level time-aware attention module and a prediction layer. As the relative future sequence for current user is actually the history sequence for the neighbors, the multi-trend routing algorithm is applied on both the future and history sequences using shared parameters in parallel. The framework takes the user historical interactions set ${X}$ as input. We use ${X^{u}_{1,N-K}}$ and ${X^{u}_{N-K+1,N}}$ to represent training and testing data of interactions sequence of user ${u}$ respectively. ${N}$ and ${K}$ denotes the selected total length of interaction sequence of each user ${u}$ and the length of training sequence respectively. For micro-videos from the set of ${X^{u}_{1,N-K}}$, embeddings are presented as ${\overrightarrow{e}_{X^{u}_{1,N-K}}}$.

The implicit user network module constructs neighbors set for each user by selecting the users that have similar micro-video preference as indicated in their past behaviors, and then extract the relative future sequences from each neighbor. The query items can be selected from the user historical interaction ${I_u}$, for simplicity, we solely choose the last one in the list, which can be both efficient and effective as demonstrated in the empirical analysis. The relative future behaviors are defined as the interacted items following the query item in the chronological order, aiming at representing dynamic preference of the user. The intuition in behind is that the user tend to have similar preference trend as users with similar historical behaviors, and that the user can have diverse and dynamic trends of preferences.

The multi-trend routing module is developed to obtain the neighbor centroids according to diverse motivation behind specific interactions with the micro-videos. Then we learn future-aware diverse trends based on history and future sequence jointly. Furthermore, the future sequence evolved user representation acquired by time-aware attention layer is concatenated with the historical behavior evolved user representation to generate the dynamic user preference representation vector. Finally we compute the user's preferences over different micro-videos from the pool by the prediction decoder. Each part will be elaborated in the following sections.

\begin{figure*}
    \centering
    \includegraphics[trim=0 1cm 0.5cm 0, width=\linewidth]{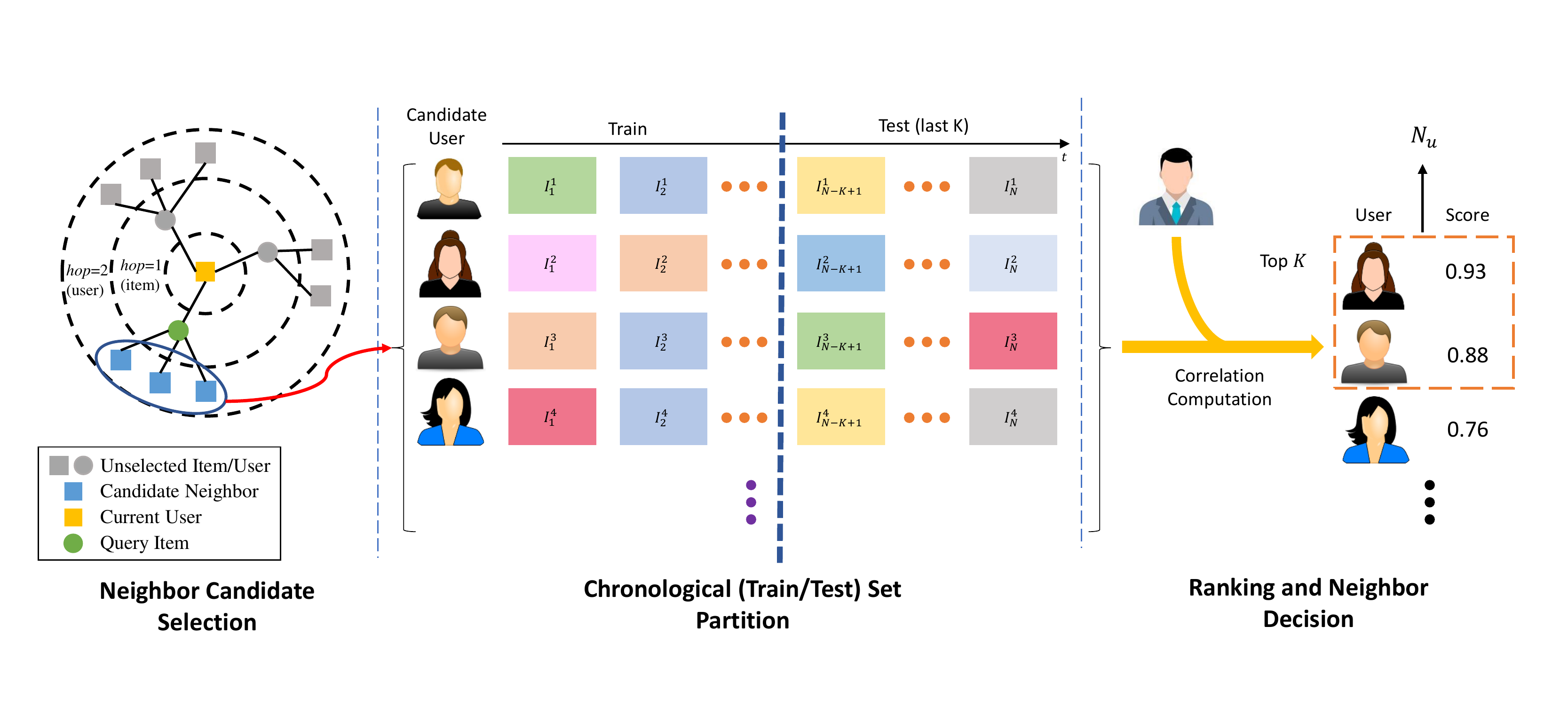}
    \caption{Architecture of the implicit user network module. The leftmost part stands for the neighbor candidate selection process based on user-item graph with the interactions by edge, user and micro-videos by node. User behaviors of the selected neighbors are then split into train and test set to compare their similarity to the current user. The relative future sequence of the most similar users are utilized to generate the future sequence as the input of multi-trend routing module, which output the future trend representation.}
    \label{fig:userItem}
\end{figure*}

\begin{algorithm}[h]
\caption{Implicit User Network Construction}
\label{algo:userNetwork}
\begin{algorithmic}[1]

\REQUIRE ~~\\ 
The set of users ${U}$;\\
User's historical interaction sequence ${I_u}$;\\
User's query items sequence ${K_u}$ and upper bound k;\\
User's candidate neighbors ${G_u}$ and upper bound g;\\
Similarity threshold ${\tau}$ for neighbor selection;\\
\ENSURE ~~\\ 
The extracted neighbor set of user ${N_u, u \in U}$;

\FOR{each $u \in U$}
\STATE ${N_u \leftarrow \emptyset}$\
\FOR{each $i \in Inverse(I_u)$}
\IF{ ${|K_u| < k}$}
\STATE ${K_u \leftarrow INSERT(i)}$\
\ENDIF
\ENDFOR
\ENDFOR

\FOR{each $u \in U$}
\FOR{each $n \in U$}
\STATE ${s_{un} = USER\_SIMILARITY(u, n)}$\
\IF{ ${n \neq u \wedge |G_u| < g} \wedge s_{un} > \tau$}
\STATE ${|G_u| \leftarrow INSERT(n)}$\
\ENDIF
\ENDFOR
\STATE ${N_u \leftarrow TOP\_SIMILARITY(G_u);}$\
\ENDFOR
\RETURN {${N_u}$}
\end{algorithmic}
\end{algorithm}

\subsection{Implicit User Network}
\label{sec:userNetwork}
As shown in Figure~\ref{fig:userItem} ,the implicit user network is constructed based on user-item heterogeneous graph, which contains both the user nodes and item nodes. An edge in the graph represents the interaction between the user and the item. The weight of the edge indicates the temporal weight of each interacted item in a chronological order. The query items are selected in a multi-hop manner. The user nodes connected to the selected query items are considered as the candidate neighbor nodes of the current user. 

Inspired by some works \cite{Guo2016socialInfluence, Felicio2016socialInfo}, which extract social relationships in absence of explicit social networks \cite{mukherjee2019ghostlink} , we construct the user network from user-item correlation implicitly.

Specifically, we compare the similarity among users via collaborative filtering implicitly based on the historical interactions with micro-videos. As the Pearson Correlation Coefficient(PCC) is a widely used similarity measure, we adopt Pearson Correlation Coefficient \cite{breese2013empirical} to compute a linear correlation between the user and each candidate neighbor as:
\begin{equation}
    s_{ij} = \frac{\sum\limits_{k\in I(i) \cap I(j)}(r_{ik} - \overline r_i) \cdot (r_{jk} - \overline r_j)}{\sqrt{\sum\limits_{k\in I(i) \cap I(j)}(r_{ik} - \overline r_i)^2} \cdot \sqrt{\sum\limits_{k\in I(i) \cap I(j)}(r_{jk} - \overline r_j)^2}}
\label{eq:pearsoncc}
\end{equation}
where ${I(i)}$ is a set of micro-videos user ${i}$ interacted with, ${r_{ik}}$ and ${\overline r_i}$ represents the level (click or not click) of interaction of user ${i}$ over micro-video ${k}$ and the average level of action of user ${i}$. The user similarity $s_{i}$ is ranging from ${[-1, 1]}$, and the similarity between users ${i}$ and ${j}$ is proportional to the value according to this definition. Following \cite{Hao2013implicitSocialRec}, we employ a mapping function ${f(x) = (x+1) / 2}$ to bound the range of PCC similarities into ${[0, 1]}$.

In the case of users with only one common micro-video in history, PCC similarity gets $1$ when the users’ preferences over the common micro-video are similar and $-1$ when not, which encourages diversity of neighbors while damaging the fairness of similarity calculation. To tackle this issue, we only kept less than ${20\%}$ of such neighbor nodes to seek the balance.

In addition to the PCC method, we also design a filter with simple schema to extract similar users. For each user, if the historical interactions ${I_u}$ is split into two pieces, ${I^{u}_{1:t_1}}$ for training data, and ${I^{u}_{t_1:t_2}}$ for testing data, the item ${\hat{I}^{u}_{k}}$ is defined as the last ${k}$ micro-videos, ${k}$ could be any value less than or equal to ${|I^u|}$, while in practice ${k=1}$
can achieve good enough performance with simplicity. We extracted a list of neighbors ${N=\{n_1, n_2, ..., n_{|N|}\}}$ according to the query item. The detail of this process is described in Algorithm~\ref{algo:userNetwork}. Furthermore, we constructed the future sequence of user ${u}$ as:
\begin{equation}
    F_u = \{n_f, n_f \in I^n, TI(n_f) \geq TI(I^{u}_{|I_u| - k})\}
\label{eq:neighborFutureSeq}
\end{equation}
where Timestamp is denoted as ${TI}$ and the query item is denoted as ${I_{|I_u| - k}}$. ${I_n}$ represents the interaction set of neighbor ${n}$ 

\subsection{Multi-trend Routing}
To capture the trend information lies in both history sequence and future sequence, we devised a multi-trend routing module into a two-stage manner to generate trend represent parallelly. Specifically, we group each micro-video from both the user's historical sequence and extracted relative future sequence into diverse trends in the first stage. The micro-videos that are grouped into the same trend are considered to be similar according to users' interactions over them and their own basic features. In the second stage, the micro-videos from historical sequence and relative future sequence are utilized to generate the representation of history and future trend group in parallel.

Based on the positive historical interaction sequence ${I_{+}}$ of user ${u}$, we represent each micro-video ${x}$ in ${I_{+}}$ as an embedding vector ${\overrightarrow{x} \in \mathbb{R} ^d}$, where ${d}$ is the embedding size. And we initialize positive history trend group as ${T_u^h \in \mathbb{R} ^{s \times d}}$ for user ${u}$, where ${s}$ denotes the number of trend groups indicated from historical sequence and ${d}$ denotes the embedding dimension of each history trend. Specifically, each trend embedding is represented as ${\overrightarrow{t} \in \mathbb{R} ^d}$.

Similarly, based on the extracted future sequence ${F_{+}}$ from the implicit user network. The positive future trend group is denoted as ${T_u^f \in \mathbb{R} ^{s \times d}}$ for user ${u}$, where ${s}$ denotes the number of trend groups indicated from future sequence and ${d}$ denotes the embedding dimension of each future trend.

In order to fine-tune the representation of each trend, we apply attention mechanism over each micro-video and the initialized trend group. Given the micro-video embedding ${\overrightarrow{x} \in \mathbb{R} ^d}$ and the trend embedding ${\overrightarrow{t} \in \mathbb{R} ^d}$, we calculate the weight between the micro-video and the trend based on a co-attention memory matrix. The micro-video from the history sequence and the future sequence are put into history trend and future trend separately. As the history sequence and future sequence is processed separately, our module is capable of capturing timeliness of trends which indicates evolved user interest.





\subsection{Multi-level Time Attention Mechanism}
As for the item-level, we use the weighted sum of historical micro-video features to obtain the current micro-video representation. Finally, we get the representation of each trend by attention mechanism on each micro-video in the trend group. As for the trend-level, we utilize the time-aware attention to activate the weight of diverse trends to capture the timeliness of each trend. Specifically, the attention function takes the interaction time of item ${i}$, the interaction time of trends and trend embeddings as the query, key and value respectively. We compute the final representation of trend representation future sequence of user ${u}$ as:
\begin{equation}
    HF_u = Attention(\overrightarrow{TI_{i}}, \overrightarrow{TI_{tr}}, \overrightarrow{t_{u}}) = \overrightarrow{t_{u}} softmax(pow(\overrightarrow{TI_{i}}, \overrightarrow{TI_{tr}}))
\label{eq:timeInterval}
\end{equation}
where Attention denotes the attention function, ${TI_{i}}$ represents the interaction time of micro-video ${i}$, ${TI_{tr}}$ represents the average interaction time of micro-videos related to the trend group, ${\overrightarrow{t_{u}}}$ represents the embedding of the specific trend group.

The trend group generated from the user's historical sequence and future sequence are then eventually updated by adding the corresponding trend group in ${T_u^h}$ and ${T_u^f}$ with the aggregation of history trend and future trend representation respectively.

\subsection{Prediction}
\label{sec:prediction}
After computing the trend embeddings from activated trends through time-aware attention layer, we apply sumpooling to both history and future trend representations.

\begin{equation}
    e_u^h = sumpooling(T_u^{h_1}, ..., T_u^{h_{s}}), e_u^f = sumpooling(T_u^{f_1}, ..., T_u^{f_{s}})
\end{equation}

And then we concatenate the history trend representation vector ${e_u^h}$ and future trend representation vector ${e_u^f}$ to form a user preference embedding ${\overrightarrow{e_u}}$ as:

\begin{equation}
    \overrightarrow{e_u} = e_u^h \frown e_u^f
\end{equation}

Given a training sample ${u, i}$ with the user preference embedding ${\overrightarrow{e_u}}$ and micro-video embedding ${\overrightarrow{e_i}}$ as well as the micro-video set ${V}$, we can predict the possibility of the user interacting with the micro-video as

\begin{equation}
    p(i|U, V, I) = \frac{exp(\overrightarrow{e_u}^T \overrightarrow{e_i})}{\sum_{v \in V} exp(\overrightarrow{e_u}^T \overrightarrow{e_v})} 
\label{eq:predictFunc}
\end{equation}

In the same way, we calculate the prediction score ${P(x | H_{-})}$ based on the negative interaction sequence, which aims to maximize the distance between the new micro-video embedding and user's negative trend embeddings.

The final recommendation probability ${\hat{p}_{ij}}$ is represented by the linear combination of ${p(x |H_{+})}$ and ${p(x | H_{-})}$. And the objective function of our model is as follows:
\begin{equation}
    \mathbb{L} = - \sum\limits_{i \in \mathbb{U}}   \left (  \sum\limits_{i\in H_{+}} \log \sigma (\widehat{p}_{ui}) + \sum \limits_{i\in H_{-}} log (1 - \sigma (\hat{p}_{ui})) \right )
\end{equation}
where ${\hat{p}_{ui}}$ denotes the prediction score of micro-video ${i}$ for user ${u}$, ${\sigma}$ represents the sigmoid activation function.

\section{Experiments}
\label{sec:EXP}
\subsection{Dataset}
MicroVideo-1.7M and KuaiShou were used as micro-video benchmark datasets in our experiments. Micro-video data and user-video interaction information can be found in each of these datasets. Each micro-video is represented by its features in these two datasets, and each interaction record includes the userID, micro-video ID, visited timestamp, and whether the user clicked the video. The two datasets' statistics are shown in Table~\ref{tab:staData}. 

\begin{itemize}
\item {\verb|MicroVideo-1.7M|\cite{Chen2018TemporalHA}}: This dataset comes from real data of micro-video sharing service in China which contains 1.7 million micro-videos.
\item {\verb|KuaiShou|}: This dataset is released by the Kuaishou Competition in China MM 2018 conference.

\begin{table}[h]
\caption{Statistics of the Datasets.}
\centering
\resizebox{\columnwidth}{!}{%
\begin{tabular}{l cc cc c}
\toprule
Dataset & users & items & interactions & train &test \\
    \midrule
    MicroVideo-1.7M   & 10,986 & 1,704,880 & 12,737,619 &8,970,310 & 3,767,309\\ 
    KuaiShou          & 10,000 & 3,239,534 & 13,661,383 & 10,931,092 & 2,730,291\\ 
    \bottomrule
\end{tabular}%
}
    \label{tab:staData}
\end{table}

\end{itemize}

\subsection{Implementation Details}
We used TensorFlow on four Tesla P40 GPUs to train our model with Adam optimizer. The following are the hyper-parameters: The micro-video embedding is 512-dimensional vectors, while the user embedding is 128-dimensional vectors. The batch size was set to 32, the optimizer was Adam, the learning rate was set to 0.001, and the regularization factor was set to 0.0001.

To find the user's similar neighbors, we used the Pearson Correlation Coefficient (PCC) described earlier. In the ablation analysis, we set neighbor numbers as 5, 20, and 50. As for the  future sequences, we cut off each neighbor's at most 100 interacted micro-videos after the current user's query items.  

\subsection{Evaluation Metrics}
To compare the performance of different models,we use \textbf{Precision@N}, \textbf{Recall@N}, \textbf{F1-score@N} and \textbf{AUC}, where N is set to 50 as metrics for evaluation.
\begin{itemize}
\item {\verb|Precision|}: Number of correctly predicted positive observations divided by the total number of predicted positive observations.
\begin{equation}
    Precision@N = \frac{1}{|U|} \sum\limits_{u \in U} \frac{|\hat{I}_{u,N} \cap I_{r}|}{|I_r|}
\end{equation}
where ${\hat{I}_{u,N}}$ denotes the set of top-N recommended  micro-videos for user u and ${I_r}$ is the total recommendation list for user u.

\item {\verb|Recall|}: Number of corrected recommended micro-videos divided by the total number of all recommended micro-videos.
\begin{equation}
    Recall@N = \frac{1}{|U|} \sum\limits_{u \in U} \frac{|\hat{I}_{u,N} \cap I_{u}|}{|I_u|}
\end{equation}
where ${\hat{I}_{u,N}}$ denotes the set of top-N recommended  micro-videos for user u and ${I_u}$ is the set of testing  micro-videos for user u.
\item {\verb|F1-score|}: F1 Score is the weighted average of Precision and Recall. It's used to balance between Presicion and Recall.
\begin{equation}
    F1-score = 2*\frac{Precision * Recall}{Precision + Recall}
\end{equation}

\item {\verb|AUC|}: AUC (Area Under the ROC Curve) is used in classification analysis to determine the quality of classifiers.
\end{itemize}

\subsection{Competitors}
To validate the effectiveness of our proposed DMR framework, we conducted experiments on two publicly available real-world datasets. The comparision to other state-of-the-art micro-video recommenders are summarized in Table~\ref{tab:modPer}.
\begin{itemize}
\item {\verb|BPR|\cite{steffen2012BPR}}: Trained on pairwise items, the Bayesian personalized ranking(BPR) maximize the difference between positive and negative items of each user in Bayesian approach.
\item {\verb|LSTM|\cite{zhangyuyu2014LSTM}}: Long short-term memory(LSTM) is a sequence model. Hidden states of each unit are aggregated to form user interest representation.
\item {\verb|CNN|}: The convolutional neural network (CNN) can be utilized to generate user interest representations based on the inter- action sequence. The max pooling layer and MLP layers are used for user interest extraction and prediction.
\item {\verb|NCF|\cite{hexiangnan2017NCF}}: As a collaborative filtering based model, NCF learns user embedding and item embedding with a shallow network and a deep network, which is able to learn an arbitrary function from data.
\item {\verb|ATRank|\cite{zhou2018deep}}: ATRank is an attention-based behavior modeling framework, which can model with heterogeneous user behaviors using only the attention model. It utilizes self-attention in multiple semantic spaces to capture behaviors interactions. The model is capable of predicting all types of user actions in a multi-task manner, which shows effectiveness over the highly optimized individual models. 
\item {\verb|THACIL|\cite{Chen2018TemporalHA}}: THACIL achieved the click-through prediction for micro-videos by modeling user’s historical behaviors. The proposed recommendation algorithm characterizes both short-term and long-term correlation within user behaviors. It also profiles user interests at both coarse and fine granularities.
\item {\verb|ALPINE|\cite{liyongqi2019ALPINE}}: To intelligently route micro videos to target users, ALPINE proposed an LSTM model based on a temporal graph, which is encoded by user's historical interaction sequence. The model captures the complex and diverse interests of users via a multi-level interest modeling layer. Moreover, the model achieves better performance by utilizing true negative samles, which indicates uninterested information.
\item {\verb|MTIN|\cite{jiang2020MTIN}}: This model is a multi-scale time-aware user interest modeling framework, which learns user interests from fine-grained interest groups. MTIN incorporates the interest group routing unit to generate user interest groups based on the interaction sequence and leverages fine-grained interest groups via item-level and group-level interest extraction unit. The distilled user interest representation is used to predict the click probabilities of micro-video candidates.

\end{itemize}

\begin{table*}[h]
\centering
    \caption{Overall Performance Comparision. The model performance of our model and several state-of-the-art baselines on two public datasets: MicroVideo-1.7M and KuaiShou-Dataset. The best results are highlighted in bold.}
\resizebox{\textwidth}{!}{%
\begin{tabular}{l cc cc cc cc cc cc}
\toprule
&\multicolumn{4}{c}{\textbf{MicroVideo-1.7M}} & \multicolumn{4}{c}{\textbf{KuaiShou-Dataset}} \\
\cmidrule(lr){2-5}\cmidrule(lr){6-9}\
\textbf{Model} & AUC@50 & Precision@50 & Recall@50 & F1-score@50  & AUC@50 & Precision@50 & Recall@50 & F1-score@50\\
    \midrule
    BPR       & 0.583 & 0.241 & 0.181 & 0.206 & 0.595 & 0.290 & 0.387 & 0.331\\ 
    LSTM      & 0.641 & 0.277 & 0.205 & 0.236 & 0.731 & 0.316 & 0.420 & 0.360\\ 
    CNN       & 0.650 & 0.287 & 0.214 & 0.245 & 0.719 & 0.312 & 0.413 & 0.356\\ 
    NCF       & 0.672 & 0.316 & 0.225 & 0.262 & 0.724 & 0.320 & 0.420 & 0.364\\ 
    ATRank    & 0.660 & 0.297 & 0.221 & 0.253 & 0.722 & 0.322 & 0.426 & 0.367\\ 
    THACIL    & 0.684 & ${\mathbf{0.324}}$ & 0.234 & 0.269 & 0.727 & 0.325 & 0.429 & 0.369\\ 
    ALPINE    & 0.713 & 0.300 & 0.460 & 0.362 & 0.739 & 0.331 & 0.436 & 0.376\\ 
    MTIN      & 0.729 & 0.317 & 0.476 & 0.381 & ${\mathbf{0.752}}$ & 0.341 & ${\mathbf{0.449}}$ & ${\mathbf{0.388}}$\\ 
    \midrule
    DMR       & ${\mathbf{0.731}}$ & 0.323 & ${\mathbf{0.478}}$ & ${\mathbf{0.385}}$ & 0.742 & ${\mathbf{0.343}}$ & 0.442 & 0.386\\
    \bottomrule
\end{tabular}
}
    \label{tab:modPer}
\end{table*}

\subsection{Results}
The model performance on the two datasets is summarized in Table~\ref{tab:modPer}. We run experiments to dissect the effectiveness of our recommendation model. We compare the performance of DMR with several commonly used and state-of-the-art models: BPR, LSTM, CNN, NCF, ATRank, THACIL, ALPINE and MTIN. All these models are running on the two datasets introduced above: MicroVideo-1.7M and KuaiShou-Dataset. According to the results shown in Table~\ref{tab:modPer}, our model DMR achieve better performance on precision over KuaiShou dataset and performs better in terms of AUC, Recall and F1-score over MicroVideo-1.7M dataset.



\begin{table*}[h]
\centering
    \caption{Effect analysis of Neighbors. The model performance with different Neighbor Number setting on two datasets: MicroVideo-1.7M and KuaiShou-Dataset. The metrics are @50. Here we set Neighbor Number to 5, 20, 50.}
\resizebox{\textwidth}{!}{%
\begin{tabular}{l cc cc cc cc cc cc}
\toprule
&\multicolumn{4}{c}{\textbf{MicroVideo-1.7M}} & \multicolumn{4}{c}{\textbf{KuaiShou-Dataset}} \\
\cmidrule(lr){2-5}\cmidrule(lr){6-9}\
\textbf{Model} & AUC@50 & Precision@50 & Recall@50 & F1-score@50  & AUC@50 & Precision@50 & Recall@50 & F1-score@50\\
    \midrule
    DMR-N5     & 0.689 & 0.319  & 0.425 & 0.364  & 0.674  & 0.333  & 0.439  & 0.378  \\ 
    DMR-N20    & ${\mathbf{0.731}}$ & ${\mathbf{0.323}}$  & ${\mathbf{0.478}}$ & ${\mathbf{0.385}}$  & ${\mathbf{0.742}}$  & ${\mathbf{0.343}}$  & ${\mathbf{0.442}}$  & ${\mathbf{0.386}}$  \\ 
    DMR-N50    & 0.668 & 0.280  & 0.282 & 0.281  & 0.652  & 0.329  & 0.404  & 0.362 \\ 

    \bottomrule
\end{tabular}
}
    \label{tab:neighborNum}
\end{table*}
Table~\ref{tab:neighborNum} compares the result of different neighbor number setting of 5, 20 and 50. Considering more neighbors could result in more diversity, but too many neighbors would dilute interest trends' embedding. Our model achieves improvements on neighbor number equals 20 over 5. Besides, it shows reduction if setting neighbor number from 20 to 50. This means the number of neighbors also play a crucial part in model performance.



The computational complexity of sequence layer modeling user and neighbors is $O(knd^2)$, where $k$ denotes the number of extracted neighbors, $n$ denotes the average sequence length and $d$ denotes the dimension of item’s representation. Capsule layer's computational complexity depends on kernel size and number of trends. Average time complexity of capsule layer scales $O(nTr^2)$, where $r$ denotes kernel size of capsule layer and $T$ denotes the number of trends. 
For large-scale applications, our proposed model could reduce computational complexity by two measures: (1)encode neighbors with a momentum encoder\cite{kaiming2020momentum}.(2)adopt a light-weight Capsule network.

\subsection{Recommendation Diversity}
Aside from achieving high recommendation accuracy, diversity is also essential for the user experience. With little information of historical interactions between the users and the micro-videos, recommendation systems learned to assist users in selecting micro-videos that would be of interest to them. Recommender systems keep track of how users interacted with the micro-videos they've chosen.

Many research works \cite{Kwon2012ImprovingDiversity,boim2011diversification,noia2014propensity,prem2013EnhancingDiversity} have been undertaken to propose novel diversiﬁcation algorithms. Our proposed module can learn the diverse trends of user preference and provide recommendation with diversity. We define the individual diversity as below:

\begin{equation}
    Diversity@N = \frac{\sum_{j=1}^{N} \sum_{k=j+1}^{N} \delta(CATE(\hat{i}_{u,j}) \neq CATE(\hat{i}_{u,k}))}{N \times (N - 1) / 2}
\label{eq:diversityDef}
\end{equation}
where ${CATE}$ represents the category of the item. ${\hat{i}_{u}}$ denotes item recommended for user ${u}$, ${j}$ and ${k}$ represents the order of the recommended items. ${\delta(\cdot)}$ is an indicator function.

Table~\ref{tab:diversity} presents comparisons with THACIL and MTIN over the recommendation diversity metric on Micro-video dataset, which provides category infromation of micro-videos. We adopt the setting of six historical trend and six future trend evolved from 5 neighbors for our model. From the table, our module DMR achieve the optimum diversity metric indicating the recommendation it provide can effectively take neighbors' interests into account.

\begin{table}[h]
    \caption{Model Recommendation Diversity Comparision on Micro-video Dataset.}
\centering
\resizebox{\columnwidth}{!}{%
\begin{tabular}{l c c c }
\toprule
MicroVideo-1.7M &\textbf{THACIL} & \textbf{MTIN} & \textbf{DMR} \\
    \midrule
    Diversity@10  & 1.9112  & 1.9940  & ${\mathbf{1.9948}}$ \\ 
    Diversity@50  & 1.9104  & 1.9948  & ${\mathbf{1.9956}}$ \\ 
    Diversity@100 & 1.9436  & 1.9950  & ${\mathbf{1.9954}}$ \\ 
    \bottomrule
\end{tabular}%
}
    \label{tab:diversity}
\end{table}

\section{Conclusion and Future Work}

In this work, we propose to capture even more diverse and dynamic interests beyond those implied by the historical behaviors for micro-video recommendation. We refer to the future interest directions as trends and devise the DMR framework. DMR employ an implicit user network module to extract future sequence fragments from similar users. A mutli-trend routing module assigns these future sequences to different trend groups and updates the corresponding trending memory slot in a dynamic read-write manner. Final predictions are made based on both future evolved trends and history evolved trends with a history-future trends joint prediction module.

This work represents one of the initial attempts to explicitly capture possible interest trends for a given historical behavior sequence, especially for ranking models and micro-video recommendation. We believe that such an idea can be inspirational to future works on learning recommender systems of high diversity. In the future, though the implicit user network module does not affect serving efficiency, we would like to explore whether more efficient and effective solutions exist to boost the training since introducing information from other users might also introduce inevitable noises. Moreover, we plan to extend the multi-trend capturing idea to more applications in recommender systems and address application-specific challenges.


\bibliographystyle{ACM-Reference-Format}
\bibliography{acm-ref.bib}

\end{document}